\documentclass[preprint,review,12pt]{elsarticle}

\usepackage{amssymb}

\journal{New Astronomy}

\begin{document}

\begin{frontmatter}

\title{The CMF as provenance of the stellar IMF ?}

\author[SVA]{Anathpindika, S.}
\ead{sumedh$\_$a@iiap.res.in}
\address{Indian Institute of Astrophysics, 2$^{nd}$-Block, Koramangala, Bangalore-560034, India}
\date{Accepted 1988 December 15. Received 1988 December 14; in original form 1988 October 11}

\begin{abstract}
 In the present work we examined the hypothesis that, a core mass function (CMF), such as the one deduced for cores in the Orion molecular cloud (OMC), could possibly be the primogenitor of the stellar initial mass function (IMF). Using the rate of accretion of a protostar from its natal core as a free parameter, we demonstrate its quintessential role in determining the shape of the IMF. By varying the rate of accretion, we show that a stellar mass distribution similar to the universal IMF could possibly be generated starting from either a typical CMF such as the one for the OMC, or a uniform distribution of prestellar core masses which leads us to suggest, the apparent similarity in shapes of the CMF and the IMF is perhaps, only incidental. The apodosis of the argument being, complex physical processes leading to stellar birth are crucial in determining the final stellar masses, and consequently, the shape of stellar mass distribution. This work entails partial Monte-Carlo treatment of the problem, and starting with a randomly picked sample of cores, and on the basis of classical arguments which include protostellar feedback and cooling due to emission from warm dust, a theoretical distribution of stellar masses is derived for five realisations of the problem; the magnetic field, though, has been left out of this exercise.
\end{abstract}

\end{frontmatter}

\section{Introduction}

 While the formation of stars is now reasonably well understood, the question related to the distribution of stellar masses at birth characterised by the so called stellar initial mass function (IMF), and in particular, about the shape of the IMF, is still far from being resolved. The multiplicity of stars is also well known, in fact, nearly half the stars in the solar neighbourhood have at least one companion, of which, the M-dwarfs form a large proportion (e.g. Duquennoy \& Mayor 1991; Fischer \& Marcy 1992; Thies \& Kroupa 2007). An empirical power-law distribution for stellar masses was first suggested by Salpeter (1955), however, this distribution was defined for relatively high mass (greater then 1 M$_{\odot}$) stars only. More recent observations have revealed stellar objects with sub-solar masses, as low as as $\sim$0.02 M$_{\odot}$, the Brown-dwarfs (BDs), which also appear to have companions. In fact, the IMF for BDs suggested by Thies \& Kroupa (2008) shows a discontinuity at the hydrogen burning limit. 

Numerous authors have observationally derived the IMF for massive star-clusters such as the IC348 (Lada \& Lada 1995; Luhman \emph{et al.} 1998; Muench \emph{et al.} 2003), NGC 1333 (Aspin, Sandell \& Russell 1994; Lada \emph{et. al.} 1996), and Ophiuchus (e.g. Bontemps \emph{et al.} 2001), to name a few. The similarity in shapes of these respective system IMFs has encouraged the idea of a universal IMF. A piece-wise form of such an IMF was proposed by Kroupa (2002), and a lognormal IMF, by Chabrier (2003). The question relating to the form of the IMF raises a more fundamental issue, that of the evolution of prestellar cores, and the factors likely to affect the final distribution of stellar masses. Infrared observations of dense prestellar cores have enabled determination of core masses in a number of nearby star-forming regions, which like the stars themselves, also follow a power-law distribution (e.g. Motte, Andr{\' e} \& Neri 1998; Andr{\' e} \emph{et al.} 2004; Alves \emph{et al.} 2007; Nutter \& Ward-Thompson 2007, Enoch \emph{et al.} 2008; Rathborne \emph{et al.} 2009). The similarity between the distribution of core masses and stellar masses is suggestive of the possibility that the stellar IMF may derive its form from the former distribution, the so called core mass function (CMF) (e.g. Motte \emph{et. al.} 1998, Nutter \& Ward-Thomson 2007). 

It has recently been shown by Swift \& Williams (2008) that the IMF is fairly robust to any variations in the evolutionary history of the prestellar cores. Similarly Goodwin \emph{et al.} (2008), starting from a power-law CMF, derived the stellar IMF by assuming a core-to-star efficiency of about 15 \%. This problem has also been examined numerically by for instance, Bonnell \emph{et. al.} (2006), Clark \emph{et. al.} (2008), who have shown that gravitational fragmentation of prestellar clumps could produce a stellar mass distribution similar to the Salpeter IMF, with a turnover located at approximately the Jeans mass of the precollapse clump, and a knee at $\sim$0.6 M$_{\odot}$. The numerically derived IMF by the above authors was shown to become shallower below its knee, consistent with the universal IMF suggested by Kroupa.

In the present work we use a sample of cores with randomly picked masses to derive the stellar-mass distribution for five realisations of the problem. First, starting with a power-law CMF defined by Eqn. (1) below, we derive the stellar mass distribution for 4 cases: (a) cores in this case are initially quiescent, self-gravitating cores are allowed to cool according to a suitable cooling-law, (b) similar to (a), however, precollapse cores are additionally supported by a transonic turbulent field, and (c) as in (a), the cores are quiescent, but self-gravitating cores are not allowed to cool. Finally, (d) for a sample of cores having a uniform mass distribution, we derive a stellar mass distribution for quiescent cores while admitting a cooling scheme. This case is repeated for a higher rate of mass accretion.  The scheme employed for the purpose is outlined in \S 2 below; the results are presented and discussed in \S 3 and \S 4, and we conclude in \S 5.


\section{Generating the stellar Initial mass function }

We wish to test how crucial is the nature of the initial distribution of prestellar cores in determining the shape of the distribution of stellar masses, in particular the resemblance of the latter with the universal IMF. This hypothesis is tested using an analytic scheme based on the following three premises :
\begin{enumerate}
\item Assumption of the validity of Larson's scaling relation defined by Eqn. (2) below, down to the spatial extent of a prestellar core. Keeping in mind the smallest length-scale over which the relation is valid, (typically 0.1 pc), our sample of prestellar cores is such that the size of the smallest core in it is larger than this minimum length scale.
\item We do not include contribution due to compressional heating of the gas in a self-gravitating core, nor do we account for viscous heating of the protostellar disk. Any additional heating will only raise the mass of putative fragments and probably produce fewer fragments, causing the final distribution to move towards the high-mass end. However, it is not our aim in a demonstrative calculation such as this one, to resolve the population of sub-stellar dwarfs, a largely recondite area in the theory of star-formation.
\item Investigation of the temporal stability of the attendant circumstellar disk is beyond the scope of the proposed deduction, so we simply work out the maximum number of sub-stellar fragments likely to condense out of the disk over the time required for the protostellar envelope to be blown off via feedback. Having accounted for protostellar feedback from the first generation of protostars, the calculations were terminated on formation of the next generation of stars, where possible, out of residual gas in a core. It is reasonable to so do since most of the gas in a core will have been consumed by the two generations of stars; some gas from the core may indeed be lost via feedback.
\end{enumerate}
   \begin{figure}
   \centering
   \vspace*{14pt}
   \includegraphics[angle=270,width=10.cm]{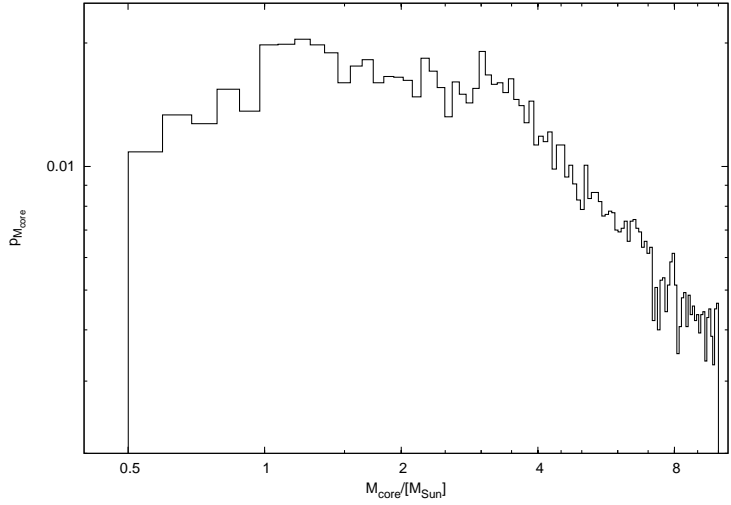}
      \caption{The histogram plotted on a logarithmic scale shows the initial distribution of prestellar cores, according to the power-law defined by Eqn. (1) below.}
         \label{FigVibStab}
   \end{figure}
%

We note that under the proposed scheme, discussed in \S 2.1, we are only estimating the maximum number of fragments likely to condense out of a self-gravitating core.
In the first three realisations of the problem, we use a sample of 14,000 randomly picked cores having individual mass, $M_{core}/\textrm{M}_{\odot}$, in the range 0.5 to 10. The core masses are distributed according to a power-law,
\begin{equation}
M_{core}\Big(\frac{dN}{dM_{core}}\Big)\propto M_{core}^{\alpha},
\end{equation} 
with slope,
\begin{displaymath}
\alpha = \left\{ 
  \begin{array}{rl}
    -1.35 &; \textrm{if } \frac{M_{core}}{\textrm{M}_{\odot}}\gtrsim 3  \\ 
    -0.3  &; \textrm{if } 1\lesssim\frac{M_{core}}{\textrm{M}_{\odot}} < 3  \\ 
     0.3  &; \textrm{if } 0.5\lesssim\frac{M_{core}}{\textrm{M}_{\odot}} < 1,  \\ 
  \end{array}\right.
\end{displaymath}
 suggested by Nutter \& Ward-Thompson (2007), for prestellar cores in the Orion Molecular cloud; $dN$ is the number of cores in the range ($M_{core}$, $M_{core}+dM_{core}$). The initial distribution of core masses in our test sample is shown in Fig. 1 above. For the purpose we used a random number seed, $\mathcal{R}(0,1)$, that produced a uniform distribution of random numbers between 0 and 1. The sample of cores within the desired range of masses is then generated by appropriately scaling the uniform distribution, $\mathcal{R}$.
The radius of a core, $R_{core}$, is calculated using the empirical length scaling relation due to Larson (1981),
\begin{equation}
R_{core} \propto \Big(\frac{M_{core}}{\textrm{M}_{\odot}}\Big)^{1/2}.
\end{equation}
 Under the approximation of a uniform density sphere for a core, the ensemble of prestellar cores in our sample has initial density typically on the order of a few times 10$^{6}$ cm$^{-3}$, modulo small variations by a factor of 1-2. The mass of putative fragments, $M_{frag}$, the number fragments, and the efficiency of fragmentation, i.e. the star-forming efficiency (SFE), were calculated for twelve choices of core temperature, $T_{core}$, in the range 7 K to 30 K. Then, in a second realisation of the problem, we raised the effective sound-speed within a precollapse core by superposing a transonic turbulent velocity field having a mean dispersion, $\sigma$, and given by the scaling relation
\begin{equation}
\sigma \propto \Big(\frac{R_{core}}{\textrm{pc}}\Big)^{0.4} \textrm{(Larson 1981)}.
\end{equation}
While acknowledging the revision to this scaling relation recently proposed by Ballesteros-Paredes \emph {et al.} (2010), we are of the opinion that it would largely serve to increase the magnitude of $\sigma$. The canonical prescription due to Larson, in our opinion therefore, is sufficient for the proposed demonstrative exploration.\\ \\
\textbf{\emph{Fragmentation of a prestellar core}}
The problem of a self-gravitating core has been studied analytically with great rigour for close to four decades, beginning the seminal work by Larson (1969), Penston (1969), followed by Hunter (1977), Shu (1979), Whitworth \emph{et al.} (1998), and Krumholz \emph{et al.}(2005). It has also been studied numerically in the recent past by Bonnell \emph{et al.}(2004), Schmeja \& Klessen (2004), Bonnell \emph{et al.} (2006), and Bonnell \& Bate (2006). These and a number of other studies have led to two suggestions in which a self-gravitating core could possibly evolve, viz. via gravitational collapse, or competitive accretion. While the former scenario envisages a self-gravitating core breaking up into a number of protostars, the latter invokes a baby protostar, typically $\lesssim$0.5 M$_{\odot}$, that grows by accreting gas from a common pool while competing with its siblings to acquire mass. The scheme adopted in this work is akin to the latter scenario, where the mass of a protostellar baby, $M_{init}$, is taken to be $\sim 0.05$ M$_{\odot}$, the Jeans mass at $\sim$7 K, the lowest choice of prestellar core temperatures in our sample.

The sound speed within a core, $a_{eff}^{2}\sim a_{0}^{2}+\sigma^{2}$, where $a_{0}^{2}=\frac{k_{B}T_{core}}{\bar{m}}$, $\bar{m}$ is the mean mass of gas molecules, for simplicity we assume the gas to be composed of molecular hydrogen and helium only; $\sigma=0$ for non-turbulent cores. We wish to emphasise that only those cores that are gravitationally bound are allowed to fragment, unbound cores are considered still-born and therefore disqualified for further calculations. We, however, do not accommodate the converse possibility of a super-Jeans core remaining starless although such cases have been reported (e.g. Sadavoy \emph{et. al.} 2010). 

Protostellar fragments in either scenarios, gravitational fragmentation or competitive accretion, grow by accreting gas, calculations by Larson-Penston and Shu arrive at an accretion rate -
\begin{equation}
\dot{M}_{acc}\propto \frac{a^{3}_{eff}}{G},
\end{equation}
which is quite different from the Bondi-Hoyle(BH) accretion rate,
\begin{displaymath}
\dot{M}_{acc(BH)}\sim\frac{\bar{\rho}G^{2}M_{frag}^{2}}{v_{g}^{3}}
\end{displaymath}
(Bondi 1952), $v_{g}$ being the terminal velocity of gas of average density, $\bar{\rho}$, accreted by the fragment; $M_{frag}$ is the mass of the fragment. Schmeja \& Klessen (2004), with their suite of numerical simulations suggested an empirical time-dependant rate of accretion 
\begin{equation}
\log(\dot{M}_{acc}) = \log(\dot{M}_{0})\frac{e}{\beta}t\exp(-t/\beta),
\end{equation}
for fitting parameters $\dot{M}_{0}$ and $\beta$. While the BH-accretion rate provides an upper limit, that due to Larson-Penston, Hunter or Shu, is constant in time. In case of a self-gravitating core, the mass, $M(r)$, within a radius $r$ is
\begin{displaymath}
M(r) = \int_{r'=0}^{r'=r_{core}}\rho(r')4\pi r'^{2}dr' = 4\pi \rho_{core}\int_{r'}\Big(r'\frac{dr'}{dt}\Big)dr',
\end{displaymath}
where the integral in the second equality on the right hand side above accounts for any modulation due to protostellar feedback. 

The foregoing discussion seeks to underline the difficulties in estimating the rate at which a protostellar fragment accretes gas. In the simplest case, that of BH accretion, $\dot{M}_{acc(BH)}\sim 10^{-4}$ M$_{\odot}$ yr$^{-1}$, provides the upper limit, while that due to Larson-Penston and Shu is $\sim 10^{-7}$ M$_{\odot}$ yr$^{-1}$, which probably, is the lower limit. Observationally determined values of $\dot{M}_{acc}$ are typically an order of magnitude higher in comparison to the latter (e.g. Bontemps \emph{et al.} 1996). In cases 1, 2, 3 and 4a, we therefore adopt a constant rate of accretion, $\dot{M}_{acc}\sim 10^{-6}$ M$_{\odot}$ yr$^{-1}$, corrected for the mass lost via feedback, as discussed in \S 2.1 below.

The mass of a protostar, $M_{frag}$, is then
\begin{equation}
M_{frag}\sim M_{init} + \frac{dM(t)}{dt}dt,
\end{equation}
where $dM(t)/dt = \dot{M}_{acc} - \dot{M}_{loss}$, $\dot{M}_{loss}$ being the rate of mass loss due to protostellar feedback. Each fragment, depending on its mass, $M_{frag}$, is assigned a temperature, $T_{frag}$, obtained from the well-known pre-main sequence evolution tracks. In the present exercise, $T_{frag}$ varies between $\sim$(1500-6500) K, for $M_{frag}$ between $\sim$(0.08-2.00) M$_{\odot}$. Feedback from a protostar ejects some mass from the natal core, and heats up the rest of it which is then tested for possible formation of the next protostar. The process is continued till the core is exhausted whence star-formation is quenched. 
 
\subsection {Feedback and cooling}
Young protostars drive energetic winds, and often, are associated with more energetic, and collimated molecular jets which inject considerable energy in star-forming regions and supposedly regulate star-formation episodes (e.g. Lada 1985). While energetic jets are likely to puncture the cocoon enclosing a protostar to deposit the associated momentum outside the parent core, the weaker protostellar wind, on the other hand, may have a significant interaction with collapsing core and heat it (e.g. Wilkin \& Stahler 1998). Below we estimate the rate at which protostellar wind drives out gas from the collapsing core with the help of a model suggested by Wilkin \& Stahler (1998). The model envisages feedback in the form a shell driven by the wind. The luminosity of the protostellar fragment as it accretes gas from the natal core is,
\begin{displaymath}
L_{acc} \sim \frac{GM_{frag}\dot{M}_{acc}}{R}.
\end{displaymath}
 The velocity, $V_{w}$, of the protostellar wind is \footnote{this equation follows from a trivial re-arrangement of the expression for energy of the driving the wind, $2E_{acc}\sim V_{w}^{2}M_{frag}\sim V_{w}^{2}\int\dot{M}_{acc}dt$.}
\begin{equation}
V_{w} \sim \Big(\frac{2L_{acc}}{\dot{M}_{acc}}\Big)^{1/2};
\end{equation}
$R\equiv R(t=t_{growth})$, being the radius of the shell. If $\dot{M}_{loss}$ is the rate of mass loss due to the wind, then a key parameter, the mass transport rate, determines if the protostellar wind could possibly escape from a self-gravitating core; it is the ratio of $\dot{M}_{loss}$ against the rate of protostellar infall, $\dot{M}_{i}$,
\begin{equation}
\alpha = \frac{\dot{M_{loss}}}{\dot{M}_{i}} \sim 0.5\Big(\frac{V_{ff}}{V_{w}}\Big)^{2}
\end{equation}
where $V_{ff}$ is the free-fall velocity \footnote{We borrow Eqns. (8) and (11) from Stahler \& Wilkin (1998); respectively, $\dot{M}_{i}\sim 4\pi R V_{w}\sigma_{shell}$, and $\sigma_{shell}\sim \frac{R^{2}\rho_{w}V_{w}^{2}}{GM_{tot}}$. Eliminating $\sigma_{shell}$ from the two equations for a spherically symmetric wind, we arrive at Eqn. (8).}; $\dot{M}_{i}\equiv dM(t)/dt$, and in the absence of feedback, $\dot{M}_{i}$ is as defined by Eqn. (4) above with 0.975 being the constant of proportionality (Shu 1977). We use Eqns. (6) and (8) to calculate the mass of a protostar. \\
\textbf{Cooling} The temperature of the envelope of gas cocooning the young protostar, and heated by it is $T_{gas}\sim T_{frag}e^{-\tau}$, $\tau$ being the optical depth within the core and $\tau\sim N_{H_{2}}s$. For a core predominantly composed of molecular hydrogen, its column density, $N_{H_{2}}$, may be estimated using an empirical relation,
\begin{displaymath}
N_{H_{2}}\sim 4.41\times 10^{21}\Big(\frac{T_{frag} a_{0}}{\mathrm{K}\cdot \mathrm{km/s}}\Big)\ \mathrm{cm}^{-2}\ \ \  \mathrm{(Tielens\  2005)},
\end{displaymath}
and a typical absorption cross-section, $s\sim 10^{-21}$ cm$^{-2}$. 
The heated gas within a self-gravitating core is allowed to cool via stimulated dust emission at infrared wavelengths. Although dust is the dominant coolant at typical protostellar densities, collisions between gas molecules and dust grains also contribute towards cooling, however, the contribution due to the latter, defined by the cooling function,
\begin{equation}
\Lambda_{gas-dust}\propto \Big(\frac{T_{gas}}{\mathrm{K}}\Big)^{0.5}(T_{gas} - T_{dust})
\end{equation}
(Goldsmith 2001), is less significant compared to the former. Cooling due to dust is quantified by the corresponding cooling function,
\begin{equation}
\Lambda_{dust} = 6.8\times 10^{-33}\Big(\frac{T_{dust}}{\mathrm{K}}\Big)^{6}\Big[\frac{n=n_{core}}{\mathrm{cm}^{-3}}\Big]\ \ \mathrm{ergs}\ \mathrm{cm}^{-3}\ \mathrm{s}^{-1}
\end{equation}
(Goldsmith 2001).

The new temperature of the core, $T_{new}$, that has radiated excess of its energy via emission from heated dust grains is,
\begin{equation}
T_{new} = \Big(\frac{(E_{shell}-E_{rad})}{4\pi r_{core}^{2}\sigma_{B}t_{ff}}\Big)^{0.25}
\end{equation}
 The residual gas in a core having acquired this new temperature, the revised Jeans mass, $M_{Jeans}(T=T_{new})$, was calculated for it. The calculations were terminated in case  $M_{Jeans}(T=T_{new})$ exceeded the available mass within the self-gravitating core; feedback from the relatively massive protostars may, however, disrupt the natal core. Heating of prestellar cores due to cosmic rays has been neglected in these calculations, for their effect is likely to be restricted to the periphery of a typical core.


\begin{figure*}
\centering
  \vspace*{14pt}
   \includegraphics[angle=270,width=10.cm]{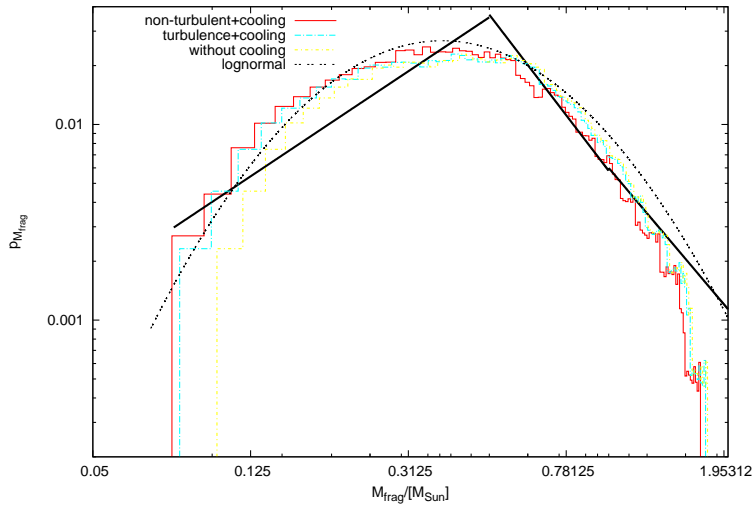}
  \caption{Starting with an initial population of cores distributed according to the power-law, see Eqn. (1) above, the derived mass distribution for fragments has been plotted for three realisations; see text for description. The universal IMF, and a lognormal fit to the derived distribution has also been plotted for comparison purposes.}
\end{figure*}

 \begin{figure}
\centering
  \vspace*{14pt}
   \includegraphics[angle=270,width=6.5cm]{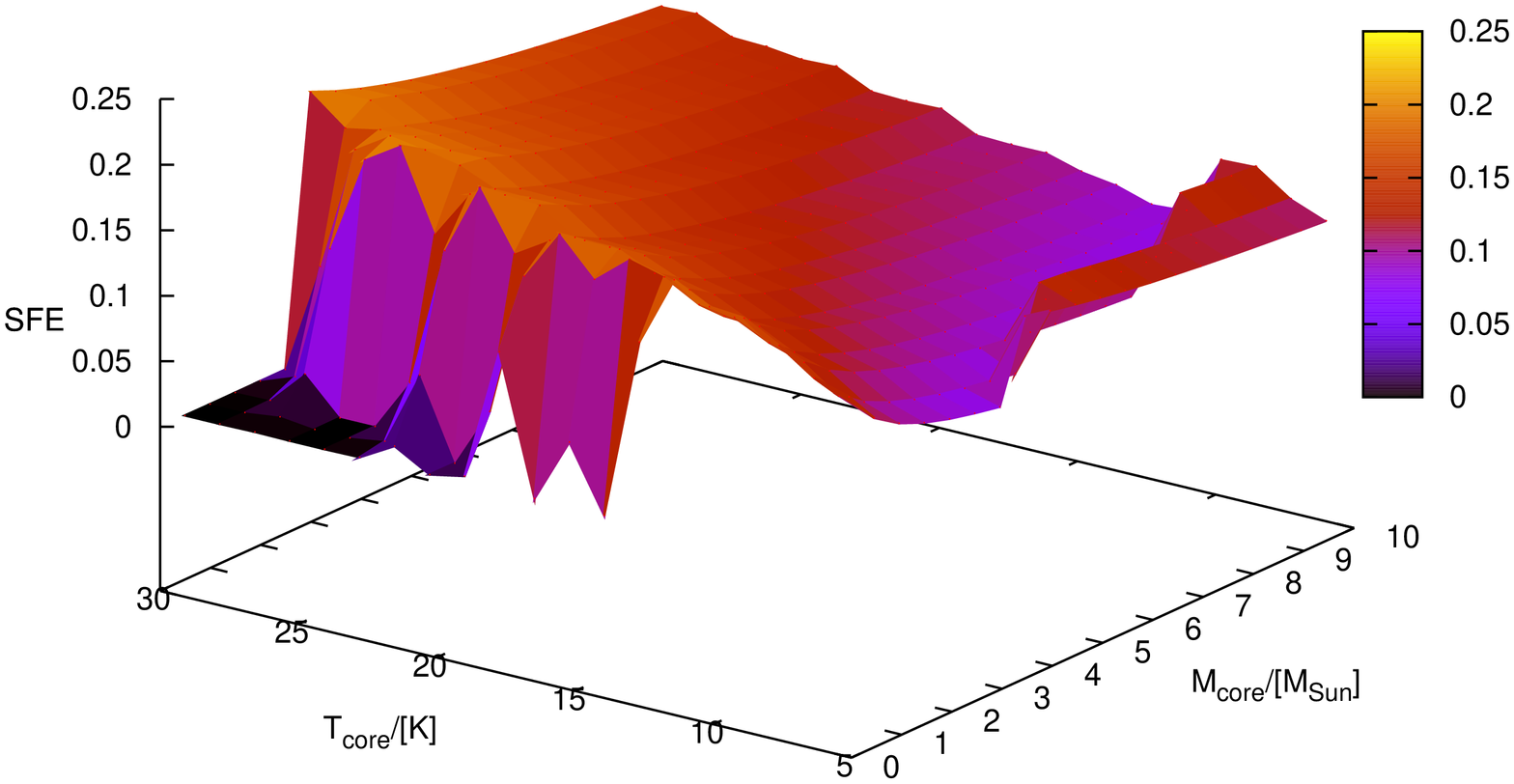}
   \includegraphics[angle=270,width=6.5cm]{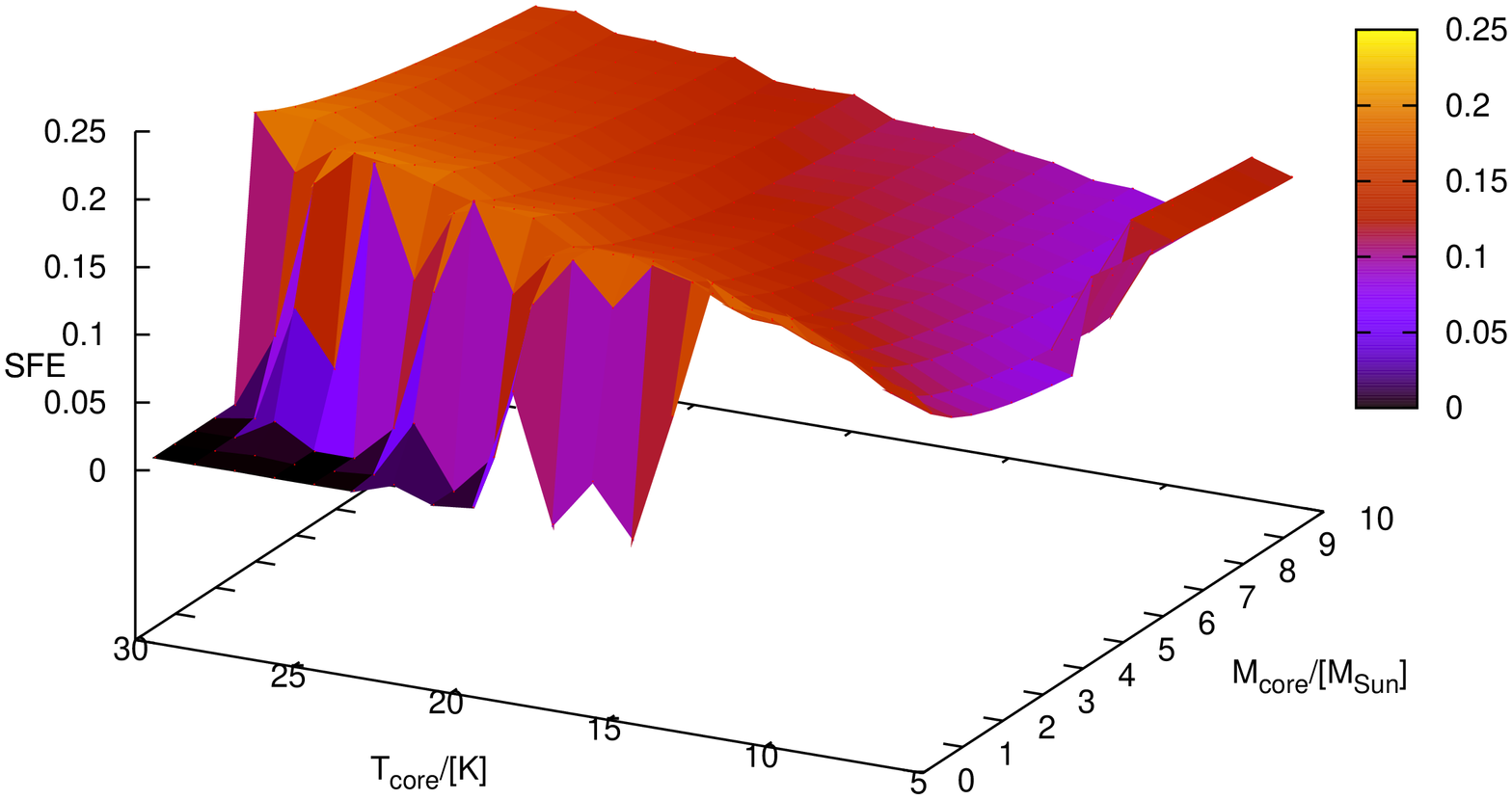}
  \caption{The efficiency of star-formation plotted collectively for all cores corresponding to the ten choices of temperature, $T_{core}$. The plot on the left is for non-turbulent cores while that on the right is for those supported by transonic turbulence. The addition of turbulence has little impact on the SFE, as is amply demonstrated by the plots. \emph{See text for description}.}
\end{figure}

\section{Results }
The relationship between the CMF and the IMF has previously been discussed on purely statistical grounds by for instance, Goodwin \emph{et al.} (2008), where the authors commencing from the CMF for the OMC suggested by Nutter \& Ward-Thompson (2007), derived the stellar IMF by randomly sampling the CMF with different choices of the core-to-star efficiency, SFE. The IMF so derived, for an SFE$\sim$ 0.15, i.e. a 15\% core-to-star conversion efficiency, matched the universal IMF suggested by Kroupa (2002). In the present work, however, having invoked physical details of a typical star-formation episode, modulo the assumptions listed above, we have adopted a different route in our attempt to deduce a distribution of stellar masses. However, our treatment of the problem is different from the hierarchical fragmentation scheme suggested by Larson (1973). While the hierarchical fragmentation scheme produced a distribution similar to the canonical IMF, the type of fragmentation envisaged in it lacks supporting numerical evidence. We shall now proceed to discuss individual realisation in this work. \\ 
\textbf{Case 1: (Non-turbulent cores)} We obtained a collective sample of $\sim$166,000 stars for various choices of core temperature, $T_{core}$, between 7K-30K. The distribution of fragment masses for this case has been plotted as a histogram in red in Fig. 2. It is characterised by a knee at $\sim$0.5 M$_{\odot}$, before steepening for masses rightward of the knee, acquiring a Salpeter-like nature. For fragments with $M_{frag}/M_{\odot}\lesssim$  0.5, the distribution becomes rather shallow and peters off steeply into the dwarf regime. The mass of the smallest fragment obtained in this case is  $M_{frag}/M_{\odot}\sim$ 0.08. While a demonstrative calculation such as the one proposed here is likely to introduce inaccuracies in the results, the distribution bears remarkable similarity to the universal IMF defined as,
\begin{equation}
dN(M_{frag})\propto M_{frag}^{-\alpha} dM_{frag}\left\{ 
     \begin{array}{ll}    
    \alpha = -1.4;0.07 \lesssim \Big(\frac{M_{frag}}{M_{\odot}}\Big)\lesssim 0.5 \\
     \alpha = 2.6;0.5 < \Big(\frac{M_{frag}}{M_{\odot}}\Big) \lesssim 1.0 \\
     \alpha = 2.3;1.0 < \Big(\frac{M_{frag}}{M_{\odot}}\Big) \\
  \end{array}\right.
\end{equation}
(Kroupa 2002), plotted using continuous bold, black segments. Also plotted, for comparative purposes is the lognormal IMF suggested by Chabrier (2003), 
\begin{equation}
N(M_{frag})\propto \textrm{exp}\Big[\frac{-(\log \textrm{M}_{frag} - \log \textrm{M}_{0})^{2}}{2\sigma_{M}^{2}}\Big].
\end{equation}
for M$_{0}\sim$ 0.08 M$_{\odot}$, and $2\sigma_{M}^{2}\sim$ 1, that also fits the red histogram, the distribution of stellar masses for this case, quite well.

Knowing the number of fragments produced by a core, the core-to-star efficiency, SFE, is calculated as-
\begin{equation}
\textrm{SFE} = \frac{\sum M_{frag}}{\sum M_{frag} + M_{core}}.
\end{equation}
The distribution of the SFE as a function of their masses and initial temperature, $T_{core}$, is shown in the left-hand panel of Fig. 3. The SFE, in general,
lies between 5\% to 15\%, while a few cores that are relatively large convert up to $\sim$25\% of their gas into stellar objects. Small, warm cores, on the other hand remain sterile as expected.  \\ 
\textbf{Case 2: (Turbulent cores)} Additional turbulent support within the precollapse cores, which in this case is transonic, not only produces fewer fragments ($\sim$ 159,000), but also shifts the distribution of fragment masses rightward, towards a higher mass, as is evident from the blue histogram shown in Fig. 2. Greater turbulent support raises the thermal Jeans mass, manifested by a rightward shift in the knee of the resultant stellar mass distribution that now appears at $\sim$0.8 M$_{\odot}$; it falls off rapidly for $M_{frag}/M_{\odot} >$0.8. The formation of dwarfs though, according to the simple arguments presented here, may not be significantly affected due to the introduction of an additional thermal component in the form of a transonic turbulent velocity field. The arguments, nevertheless bring us to the interesting question of the fate of protostellar disks in turbulent ambiance. Observational evidence in this regard is scant for lack of sensitivity of devices available at present (James d{\' i} Francesco 2010; Jenny Hatchell 2010 - private communication). One may have to wait for the commissioning of ALMA that promises to offer higher sensitivity and resolution.
Despite the additional turbulent support, the core-to-star conversion efficiency, however, does not show much change in comparison to that for case 1, apart from a slight increase in the number of cores either with a low SFE, i.e. between 5\% - 10\%, or those that remain sterile; see Fig. 3. \\ 
\textbf{Case 3: (non-turbulent cores + no cooling)}
 Cores, in this case, were allowed to fragment without invoking the cooling mechanism described in \S 2.1, although feedback from the prostellar fragments was included and ended up with even fewer fragments, $\sim 156,000$. The effect of neglecting emission from warm dust in a self-gravitating core is similar to that of adding a turbulent velocity field to cores, reported in the previous case.  The resulting distribution of fragment masses for this case is shown by the yellow histogram plotted in Fig. 2. As in the previous case, while the high-mass end, $M_{frag}/M_{\odot}\gtrsim$ 0.8, is little affected, its low-mass end shows a considerable rightward shift. There are no fragments less massive than $\sim$0.1 M$_{\odot}$. \\
\textbf{Case 4a: (non-turbulent cores + cooling; uniformly distributed core masses)} Starting with a sample of 20,000 cores with masses in the range (0.5,10)M$_{\odot}$, but now distributed uniformly, the stellar mass distribution for about 250,000 fragments derived using thermodynamic details as in case 1, has been plotted in the top panel of Fig. 4. Overlaid on top of it is the power-law,
\begin{equation}
dN(M_{frag})\propto M_{frag}^{-\alpha} dM_{frag}\left\{ 
     \begin{array}{ll}    
    \alpha = -0.7;0.08 \lesssim \Big(\frac{M_{frag}}{M_{\odot}}\Big) \\
     \alpha = -1.5;0.08 < \Big(\frac{M_{frag}}{M_{\odot}}\Big) \lesssim 0.5 \\
     \alpha = 2.35;0.5 < \Big(\frac{M_{frag}}{M_{\odot}}\Big) \\
  \end{array}\right.
\end{equation}
which is not too different from the universal IMF defined by Eqn. (12) above, making it a rather interesting case. The knee of this distribution, for instance, is still located at $M_{frag}/M_{\odot}\sim$ 0.5. It therefore leads us to the conjecture that the original nature i.e. the shape of the distribution of prestellar core masses may not be crucial to the deduction of the stellar IMF, and rather, the physical processes leading to the birth of stars are likely to determine the nature of stellar mass distribution. The often reported similarity between the CMF, and the IMF, is therefore likely to be a matter of mere coincidence or a result of the incompleteness of our surveys of young star-forming regions. 

\textbf{Case 4b} \textbf{(Same as 4a, but higher rate of accretion)} To test the role of physical processes involved, we repeated case \textbf{(4a)} with a higher rate of accretion, $\dot{M}_{acc}\sim 10^{-5}$ M$_{\odot}$ yr$^{-1}$. The resulting distribution of stellar masses is shown in the bottom panel of Fig. 4; note that the distribution is significantly overpopulated in the range (0.2,1)M$_{\odot}$. Thus, protostellar fragments accreting matter with greater efficiency tends to flatten the resulting distribution of stellar masses. The contrasting results from these two realisations re-emphasise the critical role of the physical environment in which protostars form for the rate with which gas is accreted from their natal cores depends on, both, the prevalent physical conditions within the core, and the rate at which mass is blown away by feedback.
\begin{figure*}
\centering
  \vspace*{14pt}
   \includegraphics[angle=270,width=10.cm]{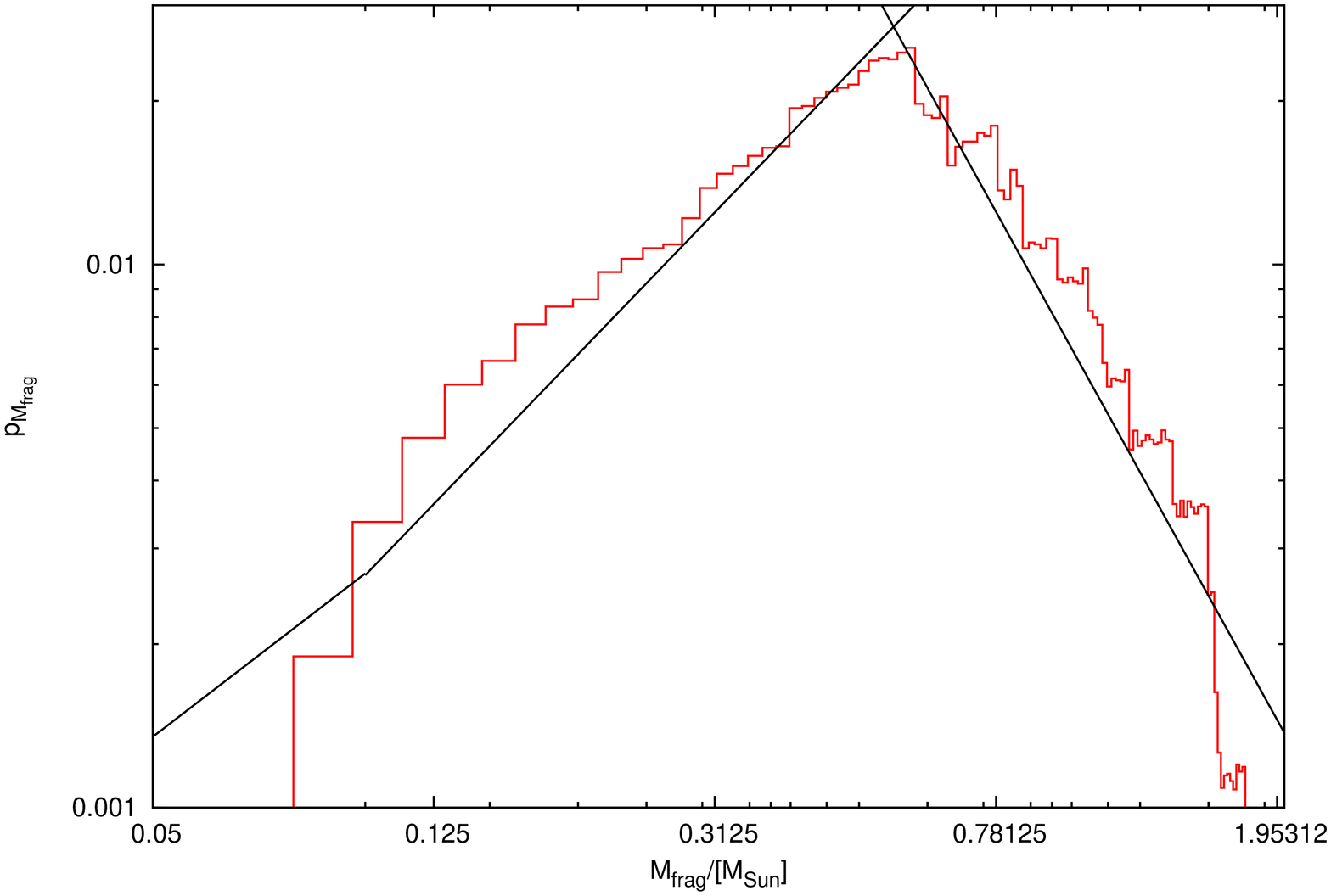}
   \includegraphics[angle=270,width=10.cm]{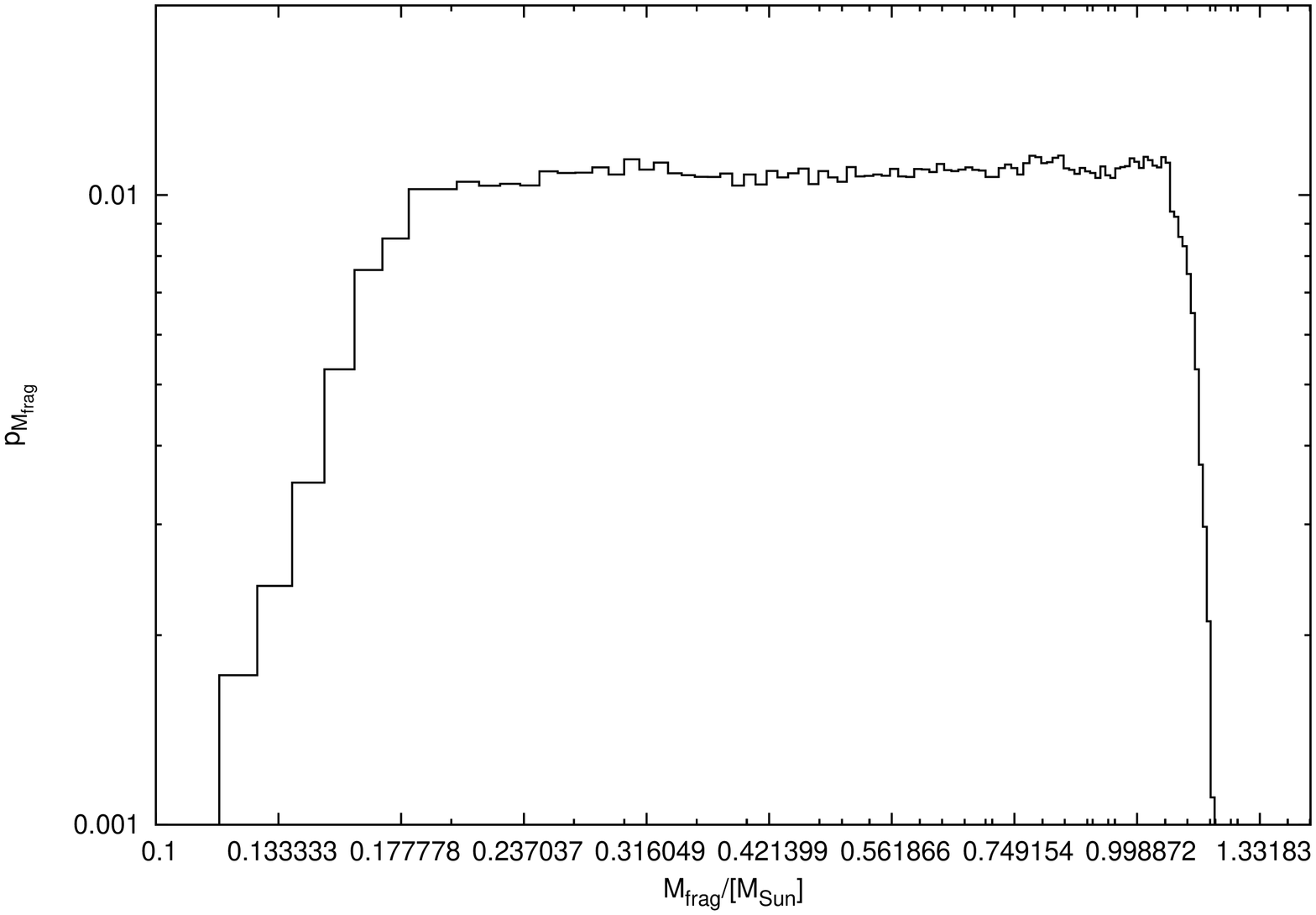}
  \caption{\emph{Top panel :}The distribution of stellar masses derived in the case commencing with a uniform distribution of core masses. The power-law given by Eqn. (15) has been overlaid for comparison purposes. \emph{Bottom panel :} The stellar mass distribution derived for conditions similar to those for the results of the realisation plotted in the panel above. The accretion rate in this case is an order of magnitude higher in comparison to the former.}
\end{figure*}

\section{Discussion}Determining the origin of the stellar IMF is a problem fundamental in nature, with arguments, both, supporting and confuting the hypothesis tested in the present work. Literary evidence favouring the hypothesis is generally encouraged by the apparent similarity between the observationally deduced CMF for a number of star-forming regions, and the stellar IMF (e.g. Motte \emph{et al.} 1998, Luhman \emph{et al.} 2006, Ward-Thompson \& Nutter 2007). On the other hand, Bonnell \emph{et al.} (2006), for instance, on the basis of their numerical simulations of self-gravitating molecular clouds argued that physical processes leading to the interplay between self-gravity and the thermal pressure, and possibly the magnetic field (e.g. Kirk \emph{et al.} 2009), could perchance, determine the nature of stellar mass distribution. In this work we examined the hypothesis in light of a simple semi-analytic scheme outlined in \S 2 above. 

The four cases discussed above seek to emphasise the importance of thermodynamics and the rate of accretion in a self-gravitating core. While an accretion rate of $\sim 10^{-6}$ M$_{\odot}$ yr$^{-1}$ in cases 1, 2, 3 and 4a led to a distribution that resembled the universal IMF, increasing it by an order of magnitude produced a flat distribution of stellar masses; see plot for case 4b in the bottom panel of Fig. 4. We therefore, as evidenced by results in case 4a, argue that a power-law CMF is not a necessary preclude to a stellar mass distribution that resembles the universal IMF.

We note that the models in cases 1, 2, and 4a which include contribution due to dust cooling and a relatively lower rate of accretion, $\dot{M}_{acc}\sim 10^{-5}$ M$_{\odot}$ yr$^{-1}$, irrespective of additional turbulent support, produce dwarfs as small as $\sim$0.08 M$_{\odot}$, however, elimination of dust-cooling from calculations in case 3 shifts the stellar mass distribution towards a 
higher mass, with a knee at 0.8 M$_{\odot}$ without any fragments smaller than $\sim$0.1 M$_{\odot}$. We allude this deficiency to the upward revision of the initial protostellar mass, $M_{init}$, due to warmer gas. Remarkably enough, starting from a uniform distribution of core masses in case 4a, but with thermal details identical to those in case 1, the stellar mass distribution derived here resembles the universal IMF which emphasises the relative importance of physical processes such as the rate of protostellar accretion over the initial distribution of core masses. Increasing $\dot{M}_{acc}$, as in case 4b, for instance, led to a considerable overpopulation in the range (0.2,1)M$_{\odot}$, rendering the distribution flat. Having demonstrated that a power-law CMF of the type defined by Eqn. (1) above need not necessarily preclude a stellar mass distribution that resembles the universal IMF, the argument favouring a direct one-to-one relationship between the two is called into question; we do not envisage one on this line either.

Cores, in general, appear to convert between 5\% to 15\% of their gas  in to stellar fragments, while a few cores of intermediate masses may have a gas-to-star conversion efficiency of up to 25 \% as can be seen from plots of the SFE in Figs. 3. The SFE for a sample of cores, as defined by Eqn. (14) above appears agnostic to changes in ambient conditions within putative star-forming cores, albeit the evolution of an individual core is governed critically by the prevalent ambient conditions. The SFE derived here is consistent with that reported for star-forming regions in the Serpens, Perseus, and the Ophiuchus molecular clouds (e.g. Enoch \emph{et al.} 2008, and references therein). Despite the success, the scheme is too naive to encompass the complexity of the star-forming process. It may also suffer owing to the lack of a complete understanding of the formation of dwarfs; possible explanations for which range from ejection at the embryonic stages (e.g Reipurth \& Clarke 2001), or via fragmentation of protoplanetary disks (e.g. Goodwin \& Whitworth 2007, Stamatellos \& Whitworth 2009).

 Finally, we must consider the possible effect of the timescale on which a core evolves, on the  distribution of stellar masses. The problem was broached by Clark \emph{et al.} (2007), the prestellar cores in the present calculations, however, are immune to this problem as they all had roughly similar initial densities, and therefore evolve on a mutually comparable timescale. In view of the empirical nature of Larson's scaling relations and their application over a large spatial range, we wish to suggest that the time-scale problem as enunciated by the above authors may not really be a nemesis to the stellar IMF.

\section{Conclusions}
For two samples of prestellar cores, distributed according to : \textbf{(A)} a power-law distribution defined by Eqn. (1) above, and \textbf{(B)} uniformly,
with masses in the range 0.5 M$_{\odot}$ to 10 M$_{\odot}$,  we have derived a distribution of stellar masses. We draw the following inferences :
\begin{enumerate}
\item By invoking a simple cooling mechanism and protostellar feedback, a stellar mass distribution similar to the IMF, is recovered for either choices of the CMF, \textbf{(A)} or \textbf{(B)}, with $\dot{M}_{acc}\sim 10^{-6}$ M$_{\odot}$ yr$^{-1}$, consistent with that inferred observationally; a higher rate of accretion, however, produces too many sub-Solar protostars as can be seen in the bottom panel of Fig. 4, and could thus, be a crucial parameter in determining the final shape of stellar mass distribution. 
\item The distribution of stellar masses appears sensitive to the thermal properties of self-gravitating cores, and the rate at which protostars accrete matter from the natal cores (Cases 1, 2 \& 3 in \S 3), interestingly enough, it appears intransitive to the preceding distribution of core masses (Cases 4a \& 4b). We therefore suggest, the apparent similarity between the observationally derived CMF, such  as the one given by Eqn. (1), and the universal IMF is possibly a coincidence. 
\item The SFE, however, appears unaffected by the injection of a little turbulence in parent cores (case 2 in \S 3), although the stellar mass distribution is shifted a little towards the higher mass end, and the shift is considerable when self-gravitating cores are not allowed to cool. There is a deficiency of fragments less massive than $\sim$0.1 M$_{\odot}$ in the latter case.
\item The prestellar cores in our test sample had roughly similar initial density modulo variation by a factor of 2-3, and so, they evolved on a mutually comparable timescale. In view of the empirical nature of the scaling relations defined by Eqns. (2) and (3) above, it appears that despite spanning a wide range of masses, the prestellar cores may have approximately comparable initial densities, and therefore, evolve over similar timescales. The distribution of stellar masses derived here is therefore fairly robust, and inoculated against variations in the evolutionary timescales of the cores in the ensemble. In order to place this argument on a stronger footing, we must further investigate the formation of prestellar cores, and their congenital properties. 
\end{enumerate}

\section*{Acknowledgements}
The author wishes to thank an anonymous referee for a critical review of the original manuscript, and acknowledges useful interactions with James d{\'i} Francesco, Derek Ward-Thompson and Steve Stahler on the matter of protostellar outflows and jets, and cooling in protostellar cores. The author is supported by a post doctoral fellowship at the Indian Institute of Astrophysics.

\label{lastpage}

\end{document}